\documentclass[
    ,final            
  ]
  {aipproc}

\layoutstyle{8x11double}

\usepackage{amsmath}
\usepackage{amssymb,amsfonts}
\usepackage{latexsym}

\def\be{\begin{equation}}
\def\ee{\end{equation}}
\newcommand{\bea}{\begin{eqnarray}}
\newcommand{\eea}{\end{eqnarray}}
\def\l{\lambda}

\def\la{\lambda}
\def\m{\mu}
\def\n{\nu}

\def\pa{\partial}
\def\t{\tau}

\def\k{\kappa}

\def\order#1{$\mathcal{O}\left(#1\right)$}

\def\hri#1#2{\href{http://arxiv.org/abs/#1}{[arXiv:#1 #2]}}
\def\hre#1#2{\href{http://arxiv.org/abs/#1/#2}{[arXiv:#1/#2]}}

\def\href#1#2{#2}

\begin{document}

\title{Recent progress in backreacted\\ bottom-up holographic QCD}



\author{Matti J\"arvinen\footnote{jarvinen@lpt.ens.fr}}{
  address={Laboratoire de Physique Th\'eorique,
\'Ecole Normale Sup\'erieure \&\\ Institut de Physique Th\'eorique Philippe Meyer,\\ 24 rue Lhomond, 75231 Paris Cedex 05,
France},
address={\centering and},
address={Crete Center for Theoretical Physics, Department of Physics,\\ University of Crete,
 71003 Heraklion, Greece}
}

%

\begin{abstract}
Recent progress in constructing holographic models for QCD is discussed, 
concentrating on the bottom-up models which implement holographically the renormalization group flow of QCD.  
The dynamics of gluons can be modeled by using a string-inspired model termed improved holographic QCD, and flavor can be added 
by introducing space filling branes 
in this model. The flavor fully backreacts to the glue in the Veneziano limit, giving rise to a class of models which are called V-QCD. 
The phase diagrams and spectra of V-QCD are in good agreement with results for QCD obtained by other methods.
\end{abstract}

\maketitle 
%
\begin{minipage}{0.97\textwidth}
\begin{flushright}
 \vspace{-9.2cm}
 CCTP-2015-02\, \\
 CCQCN-2015-60
 \vspace{9.2cm}
\par\end{flushright}
\end{minipage}

\vspace{-1.4cm}

\section{Introduction and motivation}

Holographic  QCD has been the topic of numerous studies after the discovery of gauge/gravity duality. The studied models can be roughly divided into two classes, the top-down and the bottom-up models. 

The top-down models are fixed constructions directly based on string theory and motivated by the original AdS/CFT duality. In particular, the dual field theory can usually be identified exactly, and the gravitational description does not bring in any extra parameters: all parameter of the gravitational action are uniquely fixed. The ultimate goal of this kind of constructions would be to to find an explicit holographic realization for the infrared (IR) physics of QCD, at large number of colors $N_c$. 
However, so far the available models have some shortcomings. 
Most importantly, when classical calculations are reliable on the gravity side, the low-energy spectrum of the dual field theory has additional (Kaluza-Klein) states at the same mass scale as the states of QCD. Nevertheless, the observables computed in these models are in good agreement with our knowledge of QCD (see~\cite{d3d7,wss,Erdmenger:2007cm} for 
concrete examples).

In the bottom-up approach one usually takes only the main ideas from string theory. The holographic (often five dimensional) model is constructed ``by hand'' by picking the most important operators on the field theory side, and then writing down a natural action for their dual fields. This means that the coupling constants, masses and/or potentials   of the constructed action are not known, but need to be chosen by comparing the results to the field theory. Therefore the holographic description is necessarily effective and and a wide class of gauge theories is modeled instead of a definite single theory. Typically the action is chosen such that it respects general symmetries and other features found in top-down constructions. They are able to produce a very precise description of QCD data (such as the meson spectrum) just by fitting a few parameters. Well-known examples of this type of models are~\cite{hardwall,Karch:2006pv,deTeramond:2008ht}.
Here we will discuss a specific bottom-up approach which is more strictly based on string theory than is typical for bottom-up models.

So far the majority of the literature on holographic QCD discusses flavor the 't Hooft or probe limit, where one takes $N_c \to \infty$ keeping $N_f$ and the 't Hooft coupling $g_\mathrm{YM}^2\,N_c$ fixed.
Here we will, however, also consider the backreaction of the flavor to the glue, which is present if we take the Veneziano limit 
instead. It is defined by
\begin{align}
N_c&\to\infty\ ,& \lambda&=g_\mathrm{YM}^2\,N_c \quad {\rm fixed}\ ,&\nonumber\\
 N_f&\to\infty\ ,&  x&={N_f\over N_c}\quad {\rm fixed}
\ .&
\end{align}

Backreaction means that the analysis will necessarily be technically involved. 
Taking it into account is, however, important for several reasons. First,  the number of light quarks $N_f \sim 2\ldots 3$ is comparable to the number of colors $N_c=3$ for ordinary QCD, so that the backreaction is expected to be strong. But there are also phenomena (e.g., phase transitions) which cannot be accessed in the probe limit at all. From arguments based on perturbation theory, it is clear that QCD has the so called conformal window: the theory has a nontrivial IR fixed point and chiral symmetry is intact for some $N_f$, $N_c$ with not so small $N_f/N_c$ (whereas in the 't Hooft limit $N_f/N_c \to 0$). 

More precisely, in the Veneziano limit, the leading coefficient of the QCD beta function vanishes at $x=11/2 \equiv x_\mathrm{BZ}$. As $x \to x_\mathrm{BZ}$ from below,
the theory has an perturbative IR fixed point, i.e., a fixed point at a small \order{x_\mathrm{BZ}-x} value of the coupling. But for small $x$ the IR physics is expected to be similar to ordinary QCD, with confinement and chiral symmetry breaking. Therefore a phase transition must take place at some critical value $x_c$ of the ratio $x$, which marks the lower edge of the conformal window. This transition has been called the conformal transition. 
It has been conjectured that right below the transition the IR fixed point is still visible as the theory displays  ``walking'' (nearly conformal) behavior, i.e., the coupling constant stays approximately constant for a large range of energies.

In the models discussed here, pure Yang-Mills is modeled by improved holographic QCD (IHQCD)~\cite{ihqcd}. This model is inspired by five dimensional string theory, but also contains a potential term which is not fixed a priori.
Flavor can be introduced in IHQCD by adding a pair of space-filling five dimensional D-branes~\cite{ckp}. Including full backreaction, by taking in the Veneziano limit, then leads to the V-QCD models~\cite{jk}.  Walking dynamics and the conformal transition have also been studied in closely related but simpler models called Dynamic AdS/QCD~\cite{Alvares:2012kr}. 

In these proceedings we will first discuss IHQCD and V-QCD in more detail. Then we will give an overview of the results obtained in the V-QCD models.

\section{Improved holographic QCD}

Improved holographic QCD (IHQCD)~\cite{ihqcd} is a model for the Yang-Mills theory. It is loosely based on five dimensional noncritical string theory. It is defined
in terms of the five dimensional action
\be \label{LIHQCD}
S_\mathrm{IHQCD}= M^3 N_c^2 \int d^5x  \sqrt{-\det g}\left(R-{4\over3}{
(\partial\phi)^2}+V_g(\phi)\right)
\ee
where the first term defines the five dimensional Einstein gravity and $\phi$ is the dilaton. 

We denote $\l = e^{\phi}$ and identify the 't Hooft coupling as the vacuum solution $\l(r)$, with the vacuum being the extremum of the action~\eqref{LIHQCD}. Here $r$ denotes the fifth bulk coordinate. The Ansatz for the metric is
\be
ds^2=e^{2 A(r)} (dx_{1,3}^2+dr^2)\,
\label{metric1}
\ee
in the vacuum solution. The factor $A$ is identified as the logarithm of the energy scale in field theory. In our
conventions the ultraviolet (UV) boundary lies at $r=0$, and the bulk coordinate runs from zero to infinity. The
metric will be close to the AdS near the UV.

The model deviates from the ``top-down'' derivation from five dimensional noncritical string theory as the dilaton potential $V_g$ is fixed by hand rather than derived from string theory. Its asymptotic behavior is determined using as a guideline known properties of glue dynamics in Yang-Mills theory. In the IR (i.e., large values of the coupling $\l$), potential which diverges as 
\be \label{VgIR}
 V_g\sim \l^{4 / 3}\sqrt{\log\la}\ , \qquad  (\l\to \infty)\ ,
\ee
generates confinement,
a mass gap, discrete spectrum and asymptotically linear glueball trajectories \cite{ihqcd,data}. Interestingly, this matches with the behavior predicted by string theory $V_g \sim \l^{4/3}$ up to the logarithmic correction, which is important to ensure exactly linear trajectories.

In the UV (i.e., as $\l \to 0$), 
the geometry is asymptotically AdS
and the coupling $\l$ has logarithmic dependence on $r$, similar the renormalization group (RG) flow of the coupling in Yang-Mills, when $V_g$ is analytic at $\l=0$. The holographic beta function defined by the above dictionary, $\beta = \l'(r)/A'(r)$,
then admits a Taylor expansion at $\l=0$ (with $\l$ evaluated on the vacuum solution). 
The coefficients in the Taylor expansion of $V_g$ are chosen such that the holographic beta function agrees with the perturbative beta function of Yang-Mills up to two-loops. 

At intermediate couplings
$V_g$ must be fitted to Yang-Mills data. Most precise constraints are obtained by studying the  thermodynamic functions~\cite{ihqcd2}. Even a two parameter fit gives a very good description of the lattice data~\cite{Panero:2009tv}.

\section{V-QCD}

A framework for adding flavor in holographic models has been introduced in~\cite{ckp}. It is realized through adding a pair of $D$4-$\bar D$4 branes which therefore fill the whole space in a five dimensional holographic model. The string stretching between the branes gives rise to a tachyon field, which is dual to the $\bar q q$ operator on the field theory side, and therefore controls chiral symmetry breaking. The dynamics of the tachyon $\t$ is governed by a tachyon-Dirac-Born-Infeld (TDBI) action, and taking an exponential potential, $V_f(\t) =\exp(-\t^2)$, for the tachyon. The framework has been tested in the probe approximation~\cite{ikp} with encouraging results.

The V-QCD models arise as the fusion of IHQCD with this framework for adding flavor, in the Veneziano limit where flavor fully backreacts to the glue~\cite{jk}.
The main degrees of freedom are therefore the ``dilaton'' $\l=e^\phi$ and the tachyon $\t$. The dictionary is as stated above: (the background value of) $\l$ is the 't Hooft coupling, and $\t$ is dual to $\bar q q$.

The terms in the V-QCD action which determine the vacuum solution are
\begin{align} \label{LVQCD}
S_\mathrm{VQCD}&= M^3 N_c^2 \!\int\! d^5x \sqrt{-\det g}\left(\!R-{4\over3}\frac{\left(\pa \l\right)^2}{\l^2}+V_g(\l)\!\right) & \nonumber\\ 
 &\phantom{=} - M^3 N_f N_c \int d^5x\ V_f(\l,\t)\\ \nonumber
 &\phantom{=}\times  \sqrt{-\det\left(g_{\m\n} + \k(\l)\pa_\mu\t\pa_\n\t\right)} \ .&
\end{align}
The first term is simply the IHQCD action~\eqref{LIHQCD} and describes the dynamics of the glue in the model. The second term is a generalized TDBI action, which describes the dynamics of the flavor. From the factors of $N_c$ and $N_f$ it is clear that both terms are leading in the Veneziano limit and there is indeed full backreaction.  

Notice that the form of the TDBI action is precisely known only  in the probe limit. 
Therefore we have introduced a more general Ansatz in~\eqref{LVQCD}. 
For the tachyon potential we assume the form
\be 
 V_f(\l,\t) = V_{f0}(\l)e^{-a(\l)\t^2}
\ee
where the dependence on the tachyon is the same as in the probe limit above. The potentials $V_{f0}(\l)$, $\k(\l)$, and $a(\l)$ would be power laws in the probe limit, but we have taken them to be arbitrary functions of $\l$.

In order to pin down the model, we need to specify the functions $V_g(\l)$, $V_{f0}(\l)$, $\k(\l)$, and $a(\l)$. It turns out that the dilaton potential $V_g(\l)$ must obey the same asymptotic constraints which were found in IHQCD. For the other functions similar constraints are found, both in the IR and in the UV.

In the IR ($\l \to \infty$), the functions are constrained, among other things, by the asymptotics of the spectra at large excitation numbers, and by requiring that the geometry has a ``good'' kind of IR singularity that uniquely fixes all IR boundary conditions. Consequently, we find the following asymptotics~\cite{Arean:2013tja}:
\be
 a(\l) \sim \mathrm{const.}\ ; \quad \k(\l) \sim \l^{-4/3}\sqrt{\log \l}\ ,  \qquad (\l \to \infty) \ .
\ee
Interestingly, these asymptotics again agree with the string theory prediction, up to the logarithmic correction in $\k(\l)$, in analogy to what was found for $V_g(\l)$ in~\eqref{VgIR}. 

In the UV ($\l \to 0$) the potentials must approach constant values in order for the dimension of the $\bar q q$ operator to be correctly reproduced, and for the geometry to be asymptotically AdS in the UV. 
In analogy to the approach taken in IHQCD~\cite{ihqcd}, we take the potentials to be analytic at $\l=0$ and match the coefficients of their expansions with those of the perturbative beta function of QCD and the anomalous dimension of the quark mass $\gamma = -d\log m_q/d \log \m$. 
This matching is done in order to guarantee correct boundary conditions for the IR dynamics, which can be more reliably modeled by holography and is therefore more interesting in this model.

The remaining freedom is the dependence of the potentials on $\l$ at intermediate values, which can be fitted to lattice and QCD data (neglecting the $1/N_c$ corrections arising when one moves from the Veneziano limit to finite $N_c$) in order to build a concrete model for QCD. Such a fit will be discussed in a future publication. The results below are based on potentials which obey the asymptotics found above, but were not yet fitted to any data.

\section{Properties of V-QCD}

\begin{figure}[!tb]
\centering
\includegraphics[width=0.45\textwidth]{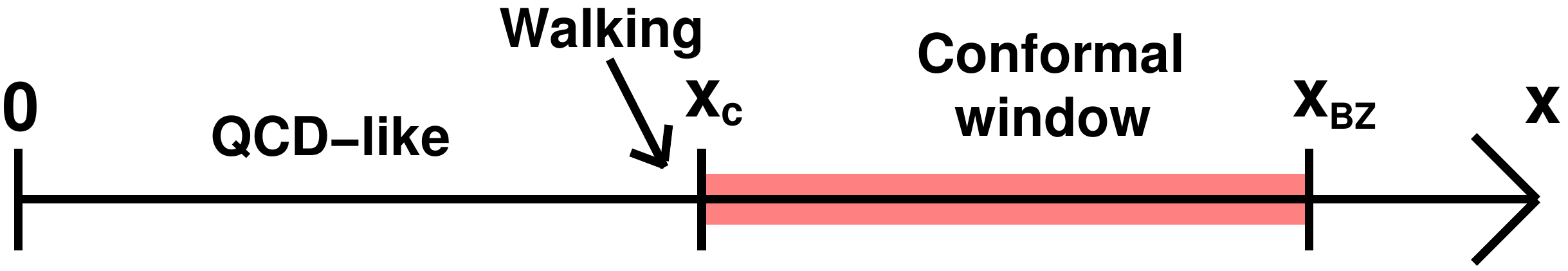}
\caption{The phase diagram of V-QCD (at zero quark mass and temperature) as a function of $x=N_f/N_c$.}
\label{fig:pd}
\end{figure}

In order to uncover the phase structure of V-QCD, we write an Ansatz where the fields $\l(r)$ and $\t(r)$ depend on the bulk coordinate $r$, and the metric has the form~\eqref{metric1}. The Ansatz is then inserted to the equations of motion derived from the action~\eqref{LVQCD}, which are consequently solved numerically.

This procedure results in phase
diagram (at zero temperature and quark mass) that is essentially universal and shown schematically in Fig.~\ref{fig:pd}. The diagram 
has two phases separated by a phase transition at $x=x_c\simeq 4$:
\begin{itemize}
 \item In the region
 $0<x<x_c$, the theory has chiral symmetry breaking, and IR dynamics is similar to that of ordinary QCD. The corresponding background solution
 has nontrivial $\l(r)$, $A(r)$ and $\tau(r)$, with the tachyon  diverging at the ``good'' IR singularity of the geometry.
 \item In the conformal window, i.e., when $x_c<x<{11/2}$, the theory flows to a nontrivial IR fixed point and is
 chirally symmetric.
  The background solution has zero tachyon $\tau(r)$=0 and nontrivial $\l(r)$ and $A(r)$, giving rise to a
 geometry flowing to a nontrivial AdS fixed point in the IR~\cite{Jarvinen:2009fe}.
\end{itemize}
As $x \to x_c$ from below, ``walking'' RG flow of the coupling constant is found.

\begin{figure}[!tb]
\centering
\includegraphics[width=0.44\textwidth]{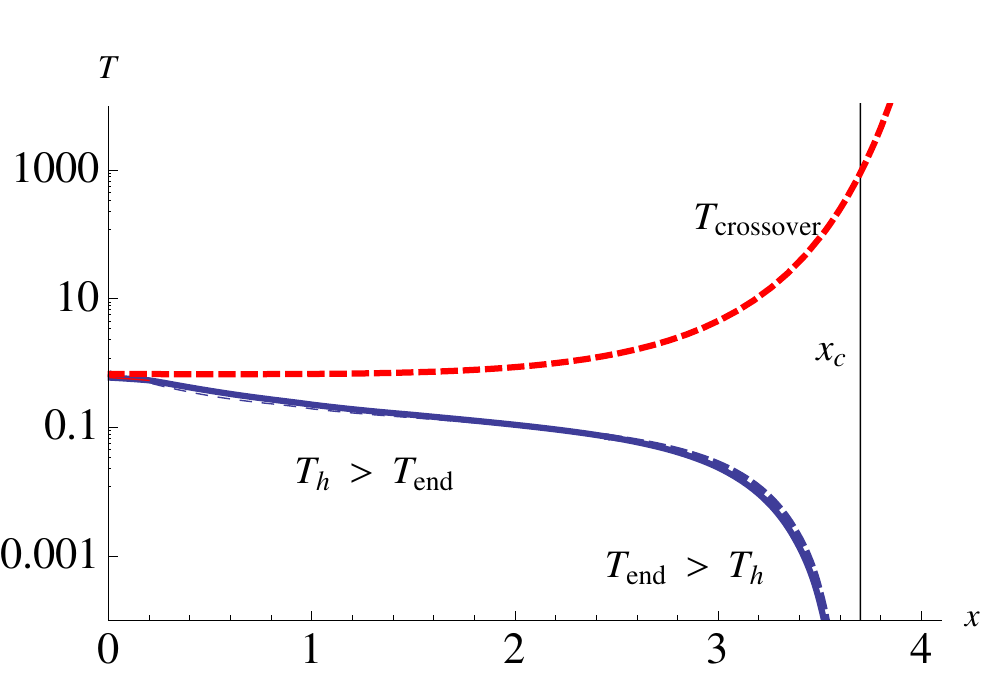}\hspace{12mm}%
\includegraphics[width=0.44\textwidth]{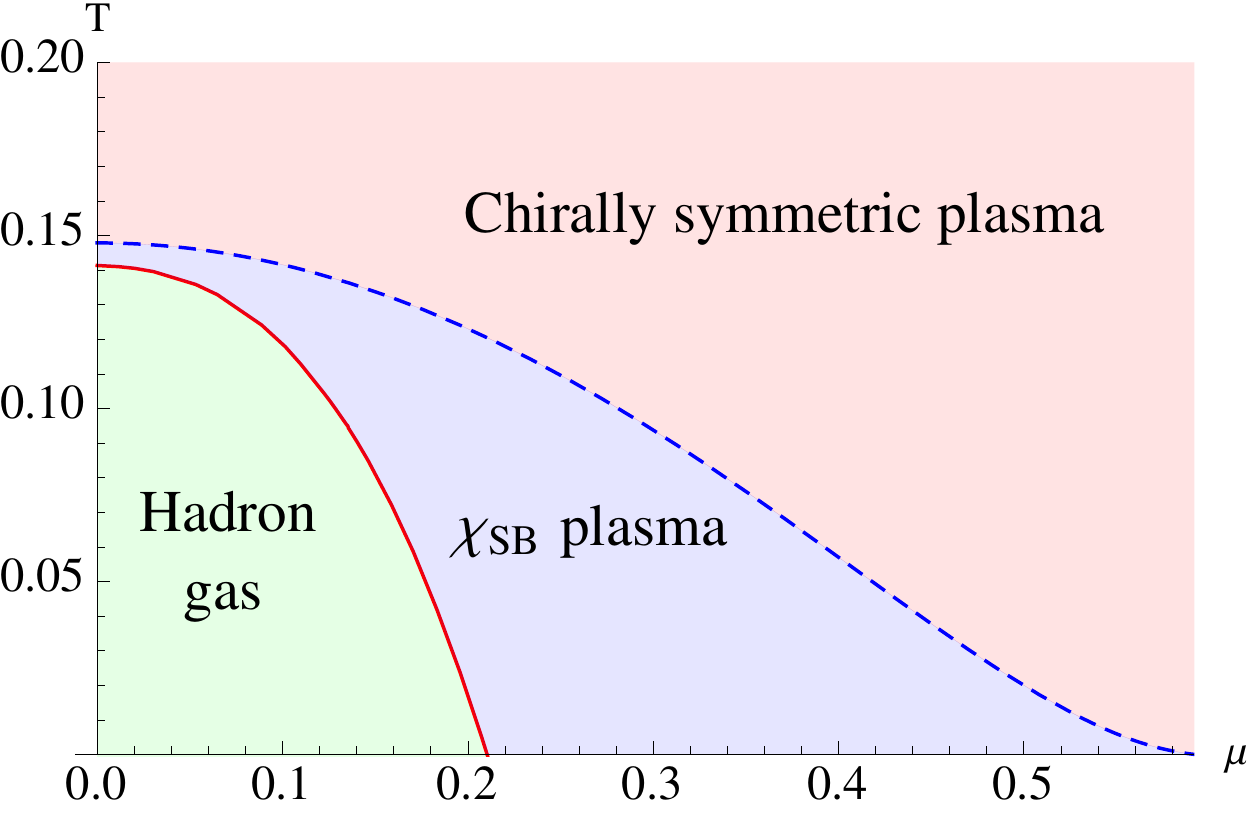}
\caption{Examples of phase diagrams of V-QCD at finite temperature on the ($x,T$)-plane (left) and at finite chemical potential, on the ($T,\m$)-plane with fixed $x=1$ (right) ~\cite{alho,Alho:2013hsa}.}
\label{fig:pdFTmu}
\end{figure}

The location of the Banks-Zaks region near $x = 11/2$, and the existence of the perturbative fixed point is a consequence of matching the potentials to the perturbative RG flow of QCD in the UV. The conformal transition and the walking regime, on the contrary, are generated by the dynamics of the model in a nontrivial way~\cite{kutasov,kutasovdbi}. One can show that the conformal transition is related to a well-known instability of the tachyon field in the conformal window. This instability arises when the five dimensional bulk mass of the tachyon, evaluated at the IR fixed point, violates the so-called Breitenlohner-Freedman bound~\cite{son}. Transition triggered by the violation of this bound quite in general leads to walking and a BKT/Miransky phase transition of infinite order. This transition involves a characteristic exponential scaling law. E.g., the chiral condensate vanishes as
\be \label{condscaling}
 \langle \bar qq\rangle \sim \exp\left(-\frac{2 K}{\sqrt{x_c-x}}\right)\ , \qquad (x \to x_c{}^-) \ .
\ee

\subsection{Phase structure at finite temperature and chemical potential}

The thermodynamics of the model can be studied as for IHQCD,
by looking for solutions with a horizon, i.e., black holes.
Rich structure of black holes, with one or two scalar hairs, was found in \cite{alho}.
A typical phase diagram at zero quark mass on the ($x,T$)-plane is depicted in Fig.~\ref{fig:pdFTmu} (left). 
For low values of $x$, there is single first order deconfinement/chiral transition (solid blue curve). Near the walking regime, the chiral restoration happens at a separate second order transition (dashed blue curve). The crossover (dashed red curve) reflects the flow from the nearly conformal (walking) regime to the UV (running) regime.
The transition temperatures obey Miransky scaling as $x \to x_c$.

\begin{figure}[!tb]
\centering
\includegraphics[width=0.44\textwidth]{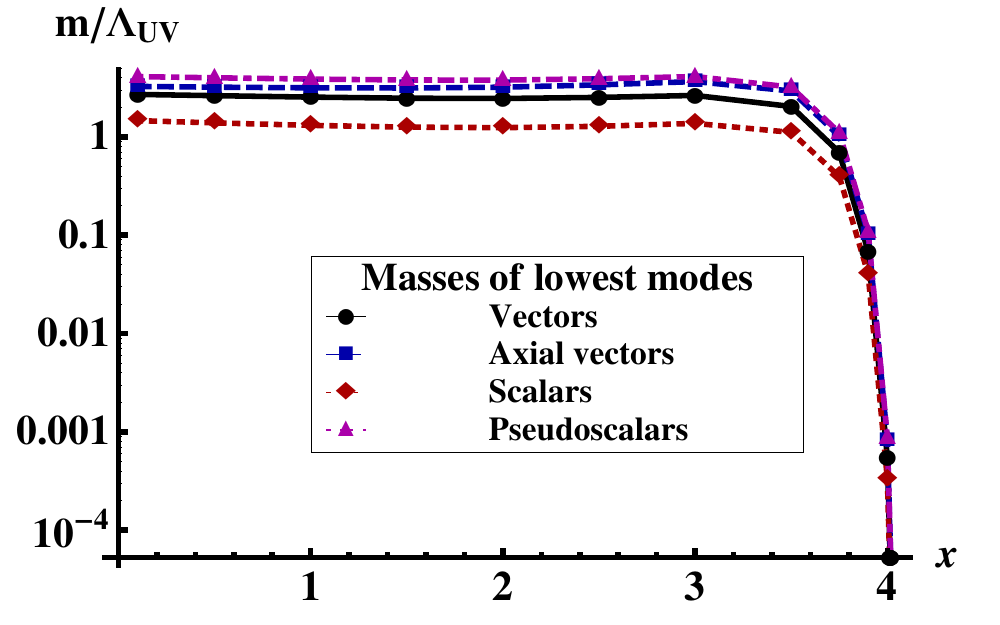}\hspace{12mm}%
\includegraphics[width=0.44\textwidth]{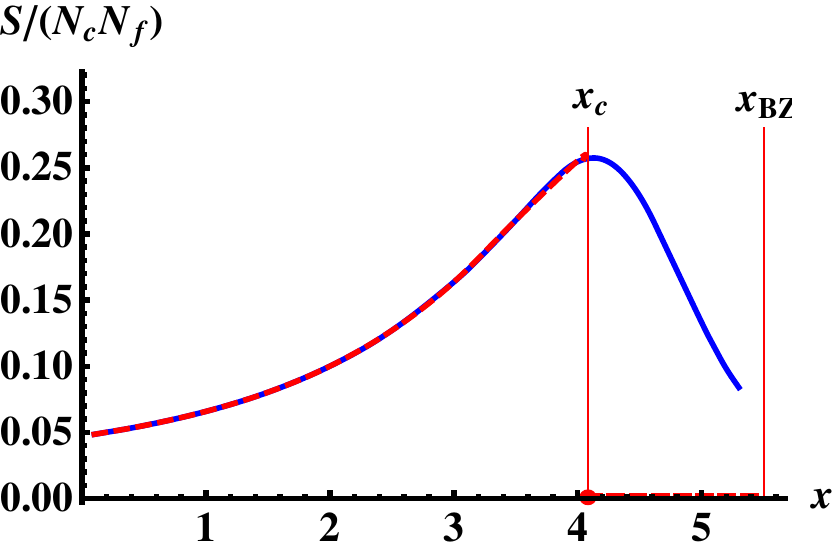}
\caption{Left: The flavor singlet masses as a function of $x$ in V-QCD. The mass of the lowest state at each sector is shown. Right: The S-parameter as a function of $x$ in V-QCD at $m_q=0$  (the dashed red curve) and at $m_q=10^{-6}$ (the solid blue curve). }
\label{fig:massesSparam}
\end{figure}

Quark chemical potential can be added in the model following the standard recipe, i.e., by allowing for a nontrivial gauge field $A_0(r)$ in the TDBI action in~\eqref{LVQCD}, and by setting $\mu =A_0(0)$~\cite{Alho:2013hsa}. An example of the resulting phase diagram is shown in Fig.~\ref{fig:pdFTmu} (right), where $x=1$. The red solid curve is the first order deconfinement transition, and the blue dashed curve is the second order chiral transition. As $T \to 0$ in the chirally symmetric phase, the IR geometry is asymptotically $AdS_2 \times \mathbb{R}^3$ and the entropy approaches a finite value instead of zero. These observations suggest that there is an instability, which may be linked to the color flavor locked or color superconducting phases of QCD.

\subsection{Spectra and the S-parameter}

The fluctuations around the vacuum solution of the V-QCD action have been studied at zero temperature~\cite{letter,Arean:2013tja} and at finite temperature~\cite{Iatrakis:2014txa}. Full analysis requires including flavor structure and gauge fields in the TDBI action (see~\cite{Arean:2013tja} for details). There are both flavor singlet (corresponding to quark bilinear operators contracted with the unit matrix $\mathbf{1}_{N_f\times N_f}$ in flavor space) and nonsinglet (corresponding to operators with the traceless Hermitean generators $t^a$) fluctuations. The normalizable fluctuation modes are identified with meson states. 

The masses of the nonsinglet mesons are relatively easy to compute. We show the masses of the lowest mode in each sector (scalars, pseudoscalars, vectors, and axial) as a function of $x$ in Fig.~\ref{fig:massesSparam} (left). The dependence on $x$ is weak for small $x$, but in the walking regime all masses go to zero obeying the Miransky scaling law. 

There is nontrivial mixing between mesons and glueballs in the flavor singlet scalar sector. It has been conjectured that the scalar singlet state contains a ``dilaton'' as $x \to x_c$, an anomalously light Goldstone mode due to the breaking of conformal symmetry. No such state was found in the analysis of the scalar singlet states of V-QCD.

Another important quantity which can be determined from the fluctuations is the S-parameter. 
For a strongly interacting theory to be a viable candidate for the realization of technicolor, i.e., in order for the theory to pass the precision tests at LEP, this parameter must be much smaller than one. The common lore is that this parameter is suppressed in the walking regime (see, e.g.,~\cite{SinDS}), 
suggesting that walking technicolor theories are viable. 

The (normalized) S-parameter in V-QCD is shown in Fig.~\ref{fig:massesSparam} (right). The blue solid (red dashed) curves have $m_q=10^{-6}$ ($m_q=0$), respectively~\cite{letter,Arean:2013tja,mj}. We notice that the normalized S-parameter is enhanced rather than suppressed in the walking region (as $x \to x_c$ from below). It will be interesting to see if this behavior remains true after the potentials of the V-QCD action have been fitted to QCD data. 
Our results show that as a tiny quark mass is turned on in the conformal window, the S-parameter immediately jumps to a finite \order{N_fN_c} value. Such a discontinuity is consistent with recent field theory analysis~\cite{sannino}.

\section{Conclusions and outlook}

In these proceedings we have discussed the construction of V-QCD and the most important results obtained with the model so far.
This model takes into account the backreaction of the flavor degrees of freedom and therefore allows us to analyze the spectra as a function of
$x=N_f/N_c$. The results are encouraging: the model mostly agrees, at qualitative level, with earlier literature on QCD.

Several additional extensions and applications of the model are under study. These include
\begin{itemize}
 \item A study of the CP-odd sector, including the theta angle and axial anomaly of QCD.
 \item An analysis of the dependence of meson masses and correlators on the quark mass.
 \item Comparison of the thermodynamics to that of low temperature hadron gas.
 \item Fitting the model to QCD data, for example Yang-Mills thermodynamics from the lattice, experimental values of the QCD meson masses, and the equation of state at finite $x$ from the lattice.
\end{itemize}
 
\subsection*{Acknowledgments}

This work was partially supported by European Union's Seventh Framework Programme
under grant agreements (FP7-REGPOT-2012-2013-1) No 316165, PIF-GA-2011-300984, the ERC Advanced Grant BSMOXFORD 228169, the EU program
Thales MIS 375734. It was also co-financed by the European Union (European Social
Fund, ESF) and Greek national funds through the Operational Program ``Education
and Lifelong Learning'' of the National Strategic Reference Framework (NSRF)
under ``Funding of proposals that have received a positive evaluation in the 3rd and
4th Call of ERC Grant Schemes'', as well as under the action ``ARISTEIA II''. 

%


%
\bibliographystyle{aipproc}   
%
%

\end{document}